\documentclass{statsoc}

\usepackage[a4paper]{geometry}
\usepackage{graphicx}
\usepackage[textwidth=8em,textsize=small]{todonotes}
\usepackage{amsmath}
\usepackage{bbm}
\usepackage{natbib}
\usepackage{comment}
\usepackage{bm}
\usepackage{booktabs}
\usepackage{hyperref}

\usepackage{comment}

\title[Assessing present and future risk of water damage]{Assessing present and future risk of water damage using building attributes, meteorology and topography}
\author{Claudio Heinrich-Mertsching}
\address{Norwegian Computing Center,
Oslo,
Norway.}
\email{claudio@nr.no}
\author{Jens Christian Wahl}
\address{Norwegian Computing Center,
Oslo,
Norway.}
\author{Alba Ordo\~ nez}
\address{Norwegian Computing Center,
Oslo,
Norway.}
\author{Marita Stien}
\address{Gjensidige Forsikring ASA,
Oslo,
Norway.}
\author{John Elvsborg}
\address{Gjensidige Forsikring ASA,
Oslo,
Norway.}
\author{Ola Haug}
\address{Norwegian Computing Center,
Oslo,
Norway.}
\author[C. Heinrich-Mertsching et al.]{Thordis L. Thorarinsdottir}
\address{Norwegian Computing Center,
Oslo,
Norway.}

\begin{document}

\begin{abstract}
Weather-related risk makes the insurance industry inevitably concerned with climate and climate change. Buildings hit by pluvial flooding is a key manifestation of this risk, giving rise to compensations of the induced physical damages and business interruptions.In this work, we establish a nationwide, building-specific risk score for water damage associated with pluvial flooding in Norway. We fit a generalized additive model that relates the number of water damages to a wide range of explanatory variables that can be categorized into building attributes, climatological variables and topographical characteristics. The model assigns a risk score to every location in Norway, based on local topography and climate, which is not only useful for insurance companies, but also for city planning. Combining our model with an ensemble of climate projections allows us to project the (spatially varying) impacts of climate change on the risk of pluvial flooding towards the middle and end of the 21st century.
\end{abstract}

\section{Introduction}

A recent report by the World Meteorological Organization found that floods were the most common of weather-, climate- and water-related disaster types recorded in the period 1970-2019 \citep{WMO2021}. While single events of large fluvial (river) floods can cause damages worth billions of Euros \citep{Barredo2007}, a large proportion of overall flood damages is caused by pluvial flooding---surface water flooding resulting from heavy rainfall---due to the far greater reach of these events \citep{Houston&2011,Spekkers&2011}. For instance, \cite{Houston&2011} assess that around 2 million people in the UK are at risk from pluvial flooding. Pluvial floods are commonly considered an invisible hazard, as they can strike with little warning in areas with no recent record of flooding \citep{Netzel&2021}, and the risk of pluvial flooding may increase in the future due to a combination of climate change, urbanization and lack of investment in sewer and drainage infrastructure \citep{SkougaardKaspersen&2017}. 

A building's exposure to pluvial flood risk depends on a range of factors such as the building's attributes, local weather and topography. The extent of which these factors influence the risk exposure is commonly assessed based on insurance claims data on reported flood damages, see \cite{Gradeci&2019} for a systematic review of the use of insurance claims data in analyzing pluvial flood events. A critical challenge when assessing flood impact is the lack of good quality flood impact data \citep{Hammond&2015}. One specific challenge is that insurance claims data may not separate between fluvial and pluvial flood damage \citep{Bernet&2017}. In Norway, however, fluvial flood damage is covered by a compulsory natural perils insurance linked to fire insurance and managed by the Norwegian Natural Perils Pool, while pluvial flood and other rainfall-induced damage is covered by a private insurance and managed directly by the primary insurer. In the following, we do not separate between pluvial flood and other rainfall-induced damage, and, for simplicity, we refer to these as water damages.   

The study of water damage and its relationship to meteorological, hydrological and topographical variables is nicely summarized in \cite{Lyubchich2019} and \cite{Gradeci&2019}. A commonality among these studies is that the number of claims and the claim size is aggregated in space (see Table 6 in \cite{Lyubchich2019}), for example at the level of municipality or postal code. Many papers also model daily claim frequency or severity and use meteorological and hydrological variables that are associated with the specific daily event, see for example \cite{Spekkers2014} and \cite{Haug2011}. For assessing the risk of specific buildings, or risk on an annual basis (for example, for pricing) there are two challenges to using a daily model on spatially aggregated data. The first challenge is that spatially aggregated data disguise building specific information, which makes it hard to assess the risk of a specific building at a specific location. Variables such as topographical indices that are available at a high spatial resolution will also lose their accuracy and thus potentially explanatory power if spatially aggregated. The challenge of a daily model that uses daily weather variables is that predicting future claims beyond approximately two weeks is challenging due to the high uncertainty of possible weather outcomes and a lack of skillful long-range weather predictions for this time resolution  \citep{vanStraaten&2020}. 

The goal of this study is two-fold. Our first goal is to provide an estimation of current, or near-term, water damage risk for any building or potential building site in Norway. Our second is to project potential changes in water-damage risk in a future climate. To this aim, we employ detailed topographical information at a $20 \times 20$ m resolution over Norway and a more general quarterly summary of local weather  statistics as detailed information of future weather is unlikely to be robustly projected for a future climate. Insurance data from the insurance company Gjensidige, including information on building attributes---to account for building-specific risk---and reported water damages, is available for 729,031 unique locations in Norway within the time period 2009-2021. This data cover approximately a quarter of the national market. The output of our analysis is the expected number of water damages in a given year at a given location, or the claim frequency. Here, we compare the use of a generalized additive model with either a Poisson or a negative binomial likelihood. The parametric structure of the models is such that the individual risk components related to topography, weather and building characteristics can be extracted and assessed individually or combined.

\section{Data}

Our claim-frequency model incorporates several different datasets. Claims data from the insurance company Gjensidige are combined with topography data derived from a digital elevation model and historical meteorological data provided by the Norwegian Meteorological Institute. Moreover, regional climate projections provided by the EURO-CORDEX initiative are used to project claim frequency for future climate scenarios. A brief description of these datasets follows. 

\subsection{Insurance data}
The insurance data were provided by the Norwegian insurance company Gjensidige and contain customer information from January 1, 2009 to April 23, 2021.  The dataset contains insurance information for private houses, apartments, cabins, agricultural and industrial buildings, as well as apartment complexes with a single coverage for the entire complex, located in Norway. For each insured property, we have information on the exact location, the insurance period, the value of the property, whether the building is used as a rental property and building characteristics such as size, age, type of roof and whether the building has a basement. The dataset includes 32,534 water damage claims, or an average of 0.07 claims per insured year. Overall, the buildings in the dataset have between 0 and 15 claims. 

\subsection{Topography data}

Topographical information is obtained from the Norwegian Mapping Authority's Digital Elevation Model (DEM) corresponding to the bare-Earth surface where all natural and built features are removed. The DEM is generated from data primarily collected via airborne laser-scanning equipment and organises 10 m gridded elevation data into non-overlapping 50 km square blocks. Rather than constructing one huge national dataset, we assemble blocks into regional rectangles of manageable size, where each region typically covers 1 -- 3 counties (there are a total of 11 counties in Norway). The rectangular area is chosen large enough so that it encompasses the full hydrological catchment area for every single location within any of the counties that it represents. To further facilitate processing of the regional DEM data, a coarser spatial resolution of 20 m is established by averaging the elevation of the four 10 m cells underlying every 20 m cell.

From the 20 m gridded and regionalised DEM data, three topographical indices of particular interest to precipitation-induced damage are calculated: {\it Height Above Nearest Drainage} (HAND) \citep{Nobre&2011}, which  specifies for each grid cell in the DEM the relative elevation between the cell and the nearest waterway cell (for example, river or ocean) that it drains into. One interpretation of the HAND index is that it judges the risk of a DEM cell being affected if its associated waterway overruns its banks.
The second index is the {\em slope}, $\beta$, of a DEM cell specifies the local angle of inclination in the water flow direction out of the cell and measures the capability that a grid cell has to drain water away. A similar concept that also takes into account the amount of water potentially flowing into the DEM cell, is the {\em Topographical Wetness Index} (TWI) developed by \citet{BevenKirkby1979}. This is defined as $\mbox{TWI}=\log(a/\tan(\beta))$ where $a$ is the size of the upslope contributing area and $\beta$ is the slope. Grid cells with a high TWI therefore indicate locally flat terrain or a large upslope area, both of which increase the likelihood of accumulating water.

\subsection{Meteorological data}

Historical meteorological information is derived from the gridded data product seNorge version 2018 \citep{Lussana2020}. Using statistical interpolation of station observations, seNorge contains estimates of daily near-surface air temperature and precipitation on a grid with resolution $1$ km covering all of Norway from 1957 to the present day. The seNorge dataset is used to derive climatological indices on the same resolution, defined as seasonal means of daily temperature and precipitation for the time period 1991-2020. In addition to these two climatological indices, several alternative climatological indices were considered, including high empirical quantiles of the quarterly distributions of daily temperature, daily precipitation and multi-day precipitation as well as high return period values from intensity-duration-frequency curves for daily precipitation. However, as these did not improve the predictive performance of our models and yielded less interpretable models, we have chosen to focus on the mean values, as they give us more robust climatological indices. 

\subsection{Climate projections}\label{sec:climate_projections}

\begin{table}
\end{table}

Projections of the climatological indices for a future climate are provided by the EURO-CORDEX initiative \citep{Jacob&2020}. EURO-CORDEX provides a multi-model ensemble of regional climate projections for Europe at a spatial resolution of 12 km, obtained by running a limited-area regional climate model (RCM) using the output of a global general circulation model (GCM) as boundary conditions. In addition, the RCM output has been bias-corrected using a cumulative distribution function (CDF) transformation \citep{Michelangeli&2009, Vrac&2012} or distribution-based scaling \citep[DBS;][]{Yang&2010}, using data from the regional reanalysis MESAN \citep{Haggmark&2000,Landelius&2016} or the observation-based gridded data product E-OBS \citep{Cornes&2018} as calibration data. 

\begin{table}
\caption{\label{tab:climate_data} Overview of the EURO-CORDEX climate projection ensemble used in this paper. For each ensemble member, the general circulation model (GCM), the regional climate model (RCM) and the bias-correction method (combination of method, data product and data period) is listed. }	
\centering
\fbox{%
\begin{tabular}{lll}
		\toprule
		GCM & RCM & Bias-correction method \\
		\midrule
EC-EARTH        & CCLM4-8-17     & CDFT22s-MESAN-1989-2005\\
EC-EARTH        & CCLM4-8-17     & DBS45-MESAN-1989-2010\\
EC-EARTH        & HIRHAM5        & CDFT22s-MESAN-1989-2005\\
EC-EARTH        & HIRHAM5        & DBS45-MESAN-1989-2010\\
EC-EARTH        & HIRHAM5       & CDFt-EOBS10-1971-2005\\
EC-EARTH        & RACMO22E      & CDFT22s-MESAN-1989-2005\\
EC-EARTH        & RACMO22E      & CDFt-EOBS10-1971-2005\\
EC-EARTH        & RCA4          & DBS45-MESAN-1989-2010\\
MPI-ESM-LR      & CCLM4-8-17    & CDFT22s-MESAN-1989-2005\\
MPI-ESM-LR      & CCLM4-8-17    & DBS45-MESAN-1989-2010\\
MPI-ESM-LR      & RCA4          & CDFT22s-MESAN-1989-2005\\
MPI-ESM-LR      & RCA4          & DBS45-MESAN-1989-2010\\
 	\bottomrule
	\end{tabular}}
\end{table}

Specifically, we consider 12 different combinations of GCMs, RCMs and bias-correction approaches as listed in Table~\ref{tab:climate_data}. This multi-model ensemble of climate projections includes two different GCMs (EC-EARTH \citep{Hazeleger&2012} and MPI-ESM-LR \citep{Giorgetta&2013}), four different RCMs (CCLM4-8-17, HIRHAM5, RACMO22E and RCA4, see \citet{Jacob&2014} for details), as well as three different versions of the bias-correction approaches described above. Considering such a variety of combinations helps us better account for the uncertainties associated with each layer of modeling. 
Additionally, we consider two representative carbon pathways (RCPs): RCP4.5 is an intermediate scenario where emissions peak around 2040 and decline afterwards while RCP8.5 presents a worst-case scenario where emissions continue to rise throughout the entire 21st century \citep{Jacob&2014}. 

 An ensemble of future climatological indices for two future periods, 2031-2060 and 2071-2100, are obtained by calculating the projected differences of the climatological index between the future period and the historical period 1991-2020 from each climate model. These projected differences are then added to the historical index, derived from the seNorge data. Considering differences rather than absolute projections removes potential (constant) biases of the climate models. The projected future indices are therefore derived on the high-resolution spatial scale of 1 km provided by seNorge, but for all grid cells lying within the same $12 \times 12$ km RCM grid cell the same changes apply.

\section{Methods}\label{sec:model}

\subsection{Statistical modeling framework}

The aim of the statistical modeling is to predict the number of claims, $N_i$, for contract $i$ within a year (or another time period of interest). To this aim, we model the distribution of $N_i$ as a function of various covariates that fall in four separate classes: (1) Property specific characteristics, (2) Topographical information at the property location, (3) Climatological information at the property location, and (4) A fixed effect of which county the property is located in and a random effect over municipalities to increase the flexibilty of the model.  This way, municipalities with little information get regularized towards the mean of the county, which is important since many municipalities contain only a few observations. Currently there are 11 counties and 356 municipalities in Norway. 

For the distribution of $N_i$, we compare a Poisson and a negative binomial model, denoted $N_i \sim \text{Po}(\mu_i)$ and $N_i \sim \text{NB}(\mu_i, \theta)$, respectively. While the distribution of the Poisson model is fully determined by its mean, $\mu_i$, the negative binomial model has an additional parameter $\theta$, specified by $\mathrm{Var}(N_i) = \mu_i + \mu_i^2/\theta$. In particular, the negative binomial model exhibits larger variance than the Poisson model, and the parameter $\theta$ controls for overdispersion. 

 For both models we employ a generalized additive model \citep[GAM;][]{Elements1990}. This allows for non-linear relationships between the covariates and the number of claims without needing to specify the functional form of these relationships, while still being highly interpretable. The interpretability aspect is crucial for understanding and explaining the risk structure to both potential insurance holders and decision-makers in the context of climate change adaptation. We use a logarithmic link function, such that the relationship between the expected number of claims and the available covariates is specified by 
\begin{align}\label{mu}
    \log \mu_i = \bm{z}_i^T \bm{\gamma} + \sum_{j=1}^J f_j(x_{ij}) + u_{R[i]} + \log(l_i) + \log(v_i).
\end{align}
Here, $\bm{z}_i^T$ represents the vector of categorical variables for contract $i$ with parameter vector $\bm{\gamma}$ and $f_j$ represents a smooth function modeling the effect of the $j$th continuous covariate $x_{ij}$. The variables $l_i$ and $v_i$ represent the length of the $i$th contract and the value of the building, respectively. The variable $u_{R[i]}$ is a random effect assigned to each policy $i$ belonging to municipality $R[i] \in \{1, \dots, K\}$.  We assume that the effects, $u_1, \ldots, u_K$, for all municipalities are independent and normally distributed with a common variance $\sigma_u^2$, that is, $u_k \stackrel{\text{iid}}{\sim} N(0, \sigma_u^2)$. Specifically, the random effects are expressed as a smooth spline \citep{Wood2017}. The offset, $\log(l_i)$, accounts for the increase in expected claims over time as long as the contract is active. The use of the property value as offset is convenient in insurance risk modeling, where it is more natural to model the number of claims per time and insured value, which is more closely connected to the expected payout. We additionally consider the value $v_i$ as a property-specific covariate to avoid the unreasonable assumption that $\mu_i$ grows linearly in $v_i$.

Each $f_j$ is represented as a weighted sum of basis functions,
\[
    f_j(x_{ij}) = \sum_{k=1}^K \beta_{jk} b_{k}(x_{ij}),
\]
where the $\beta_{ij}$ are unknown parameters to be estimated from the data, and $b_1, \ldots, b_K$ are known basis functions, here given by the cubic spline basis.   

\subsection{Risk assessment}\label{sec:risk_assessment}

We define the {\it risk} of the $i$th contract as $r_i := \frac{\mu_i}{l_iv_i}$, that is, the expected number of claims per year and value of insured property. The structure of the GAM facilitates that the risk $r_i$ can be decomposed into the product of partial risks corresponding to the groups of covariates described above. 
For a given contract, we first define the non-normalized partial risk over a selection of $M$ covariates as 
\[ \log \widetilde r_i^M := (\bm{z}_i^{M})^T \bm{\gamma}^M + \sum_{j\in M} f_j(x_{ij})  + u_{R[i]}\mathbbm 1\{u_{R}\in M\},\]
where $\bm{z}_i^M$ and $\bm{\gamma}^M$ denote the natural restrictions of $\bm{z}_i$ and $\bm{\gamma}$ to $M$, and $\mathbbm 1\{u_{R}\in M\} := 1$ if the random municipiality effect is part of the selection $M$, and $\mathbbm 1\{u_{R}\in M\} := 0$ otherwise. Further, we define the partial risk over the covariate group $M$ as
\begin{align}\label{eq:risk_factor}
    r_i^M := \frac{\widetilde r_i^M}{\frac{1}{N}\sum_{i = 1}^N \widetilde r_i^M},
\end{align}
where the average in the denominator is taken with respect to all contracts. Now, separating all covariates into disjoint groups $M_1,...,M_L$, we obtain the total risk decomposition
\[r_i = r_0 r_i^1\cdots r_i^L,\qquad \text{ where }\qquad r_0 := \prod_{l = 1}^L\bigg(\frac{1}{N}\sum_{i = 1}^N \widetilde r_i^M\bigg). \]

This decomposition facilitates interpretation of the results. We may, for example, consider the group of climatological covariates $M_{\text{clim}}$ and calculate the climatological risk factor $r_i^{\text{clim}}$ for a given property. Using the normalization \eqref{eq:risk_factor}, the average climatological risk equals 1, and a value above 1 indicates higher-than-average risk due to local weather. Similarly, we may compute a topographical risk factor and a building-specific risk factor. Ranking all the normalized partial risks of a specific building or location allows for recognition of its prevailing risk drivers. Furthermore, this decomposition allows for the estimation of the partial climatological and topographical risk for every location covered by our weather and topographical data, respectively, including locations without insured buildings. 

\subsection{Inference}

In order to avoid overfitting, we employ a penalized log-likelihood given by 
\begin{equation}\label{eq:logLik}
l(\bm{\phi}) - \frac{1}{2}\sum_{j=1}^J \lambda_j \bm{\beta}_j^T\bm{S}_j \bm{\beta}_j,
\end{equation} 
where $l(\bm{\phi})$ is the log-likelihood, $\bm{\phi} = (\bm{\gamma}, \bm{\beta})$ are the model parameters, $\bm{\lambda} = (\lambda_1, \ldots, \lambda_J)$ are the smoothing parameters controlling the smoothness of the $f_j$s and $\bm{S}_j$ is a matrix where the $kl$th element equals $\int b_{jk}''(x) b_{jl}''(x) \text{d}x$. The model parameters and smoothing parameters are estimated in two steps. First \eqref{eq:logLik} is optimized w.r.t $\bm{\phi}$ holding $\bm{\lambda}$ fixed using penalized iteratively reweighted least square (PIRLS). Second, the smoothing parameters are estimated using Laplace approximated restricted maximum likelihood, holding $\bm{\phi}$ fixed. For the negative binomial model, the dispersion parameter $\theta$ is estimated alongside the smoothing parameters.

All computations are performed with R version 4.0.5 \citep{rCore} using the package {\tt mgcv} version 1.8-26 \citep{Wood2011} for parameter estimation of the GAMs. As our dataset contains millions of observations and hundreds of covariates we use the methodology proposed for large datasets in \cite{Wood2017Giga}, \cite{ Wood2014} and \cite{Li2020} as implemented in the {\tt bam} function in {\tt mgcv}.

\subsection{Model evaluation}
We evaluate and compare competing models using two proper scoring rules \citep{GneitingRaftery2007}, the mean square error (MSE) and the Brier score \citep{Brier1950}, where a smaller value equals a better performance. 

The MSE is defined as 
\[
    \text{MSE} := \frac{1}{N} \sum_{i=1}^N \frac{1}{t_i} (N_i - t_i \hat{\mu}_i)^2, 
\]
where $N_i$ is the observed number of claims for the $i$th contract, $t_i$ is the length of the contract (in years) and $\hat{\mu}_i$ is the predicted expected number of claims per year. 
The scaling factor $\frac 1 {t_i}$ counteracts the fact that the variance of $N_i$ grows with the length of the contract. It can be shown that under this scaling, the rank of competing models is equal for different time units in the contract length.

The distribution of the claims data is heavily skewed. While for the vast majority of contracts we have $N_i = 0$, the data also contains many contracts with two or more claims, some contracts even reaching more than 10 claims. The MSE is sensitive to such outliers \citep{ThorarinsdottirSchuhen2018}, and an evaluation using the MSE therefore puts specific emphasis on the predictive performance for these outliers.
To complement the picture we therefore consider the Brier score as a second performance metric. Specifically, we consider the Brier score for the event of observing at least one claim,
\[\text{BS} := \frac{1}{N} \sum_{i=1}^N  (\mathbbm 1 \{N_i \geq 1\} - \hat{p}_i)^2,\]
where $\mathbbm 1$ denotes the indicator function and $\hat{p}_i$ is the predicted probability that the $i$th contract files at least one claim. This metric is insensitive to outliers and shifts the focus to predicting whether a given house will have a claim or not.

\subsection{Model selection methodology}
For assessing out-of-sample predictive performance, we perform a ten-fold cross-validation where 10\% of the contracts are removed at random during model fitting, and the fitted model is used to predict the claims for the withheld 10\%. This process is repeated ten times with each contract left out exactly once during the model training. 

We compare using either a Poisson or a negative binomial target distribution for the regression model as well as four alternative configurations of the regression equation in \eqref{mu}. The simplest baseline model includes only building-specific covariates. Additionally, we consider a model including topographical indices (TWI, slope and HAND), a model including climatological indices (mean temperature and mean precipitation) and a model including both topographical and climatological indices.  

For the climatological information, we consider both a quarterly and an annual model. For models based on quarterly climate statistics, contracts overlapping multiple quarters are split into subcontracts, one for each quarter. The claims filed under those contracts are then assigned to the subcontracts of the appropriate quarter. This increases the temporal resolution of the climate statistics, enabling the model to pick up on quarterly differences in claims connected to seasonal differences in weather. However, the splitting into quarterly subcontracts results in a different set of contracts than for an annual model, which makes it difficult to directly compare the MSE and the Brier score for these models. We address this by reversing the quarterly split before evaluation. Specifically, for the quarterly model, the $i$th contract is split into (up to) four contracts $i_1,...,i_4$ during fitting, with corresponding numbers of claims $N_{i_1},...,N_{i_4}$. After fitting the model, we obtain four different predicted means $\hat{\mu}_{i_1},...,\hat{\mu}_{i_4}$. However, for evaluation, these are not treated as four predictions, but rather we recompute $\hat{\mu}_i = \frac 1 4 (\hat{\mu}_{i_1}+\dots+\hat{\mu}_{i_4})$ and consider it as a prediction for $N_i/t_i$. Similarly, for the Brier score we obtain four probabilities $\hat{p}_{i_1},...,\hat{p}_{i_4}$ and can retrieve $\hat{p}_i$ as $\hat{p}_i = 1-\prod_{j = 1}^4(1-\hat{p}_{i_j})$, assuming independence of the number of claims across different quarters.

\section{Results}

Here, we present the results of our analysis of Norwegian water damage data. In a cross-validation study, we compare risk assessment models with different sets of covariates at two temporal resolutions and under two different distributional assumptions. We then present estimates of annual water damage risk under current and future climate. 

\subsection{Model selection}\label{sec:model selection}

\begin{figure}[t]
	\centering
	\includegraphics[width=.95\linewidth,height = 0.5\linewidth]{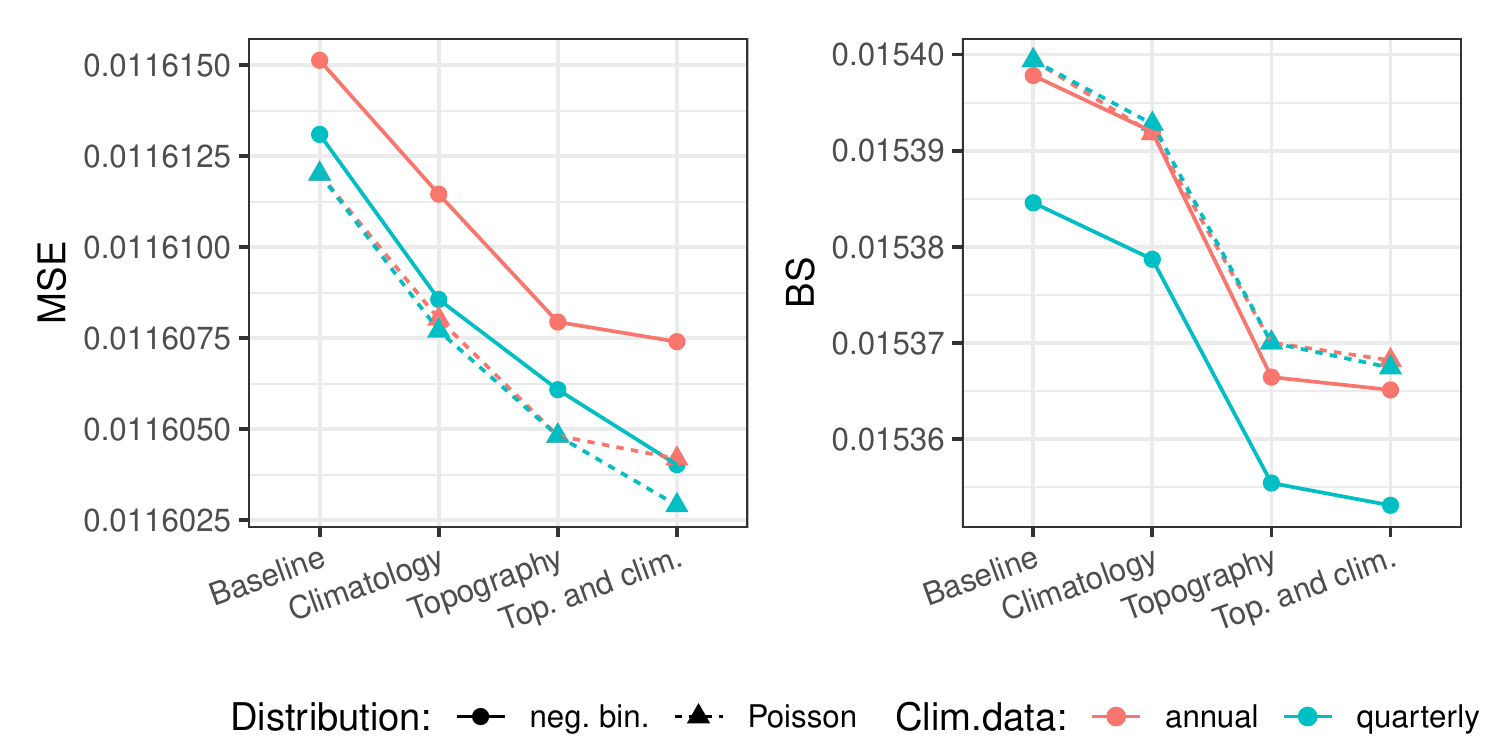}
	\caption{Comparison of models for water damage risk in Norway under a Poisson and a negative binomial likelihood with property specific covariates only (baseline), additional climatological or topographical information, or both. Annual risk is estimated based on four quarterly models, or a single annual model. We assess predicted number of annual damage claims using MSE (left) and predicted probability of seeing one of more claims per year using the Brier score (right). Scores are given as mean scores over a tenfold cross-validation of the entire dataset. }
	\label{fig:mse}
\end{figure}

Figure~\ref{fig:mse} shows the out-of-sample predictive performance of 16 different risk assessment models based on a cross-validation study, where the models differ in distributional assumptions, covariate selection and temporal resolution. Overall, a baseline model that only includes property specific information is outperformed by models that include topographical and/or climatological variables. Further, as expected, including topographical information with high spatial precision gives a greater improvement in performance than including more general climatological information. 
Under the Poisson model, the sum of the four quarterly predictions is again Poisson distributed.
The only difference is that the quarterly model resolves seasonal variation in the climatological data.

Under the negative binomial model, however, the annual-scale distributions differ for the two time resolutions. Here, the quarterly model seems significantly better than the annual one. 

While proper scoring rules favor the true data generating process by construction, they may evaluate different aspects of competing models and thus not always yield identical rankings when none of the models represent the true data generating process. The Poisson models are preferred over the negative binomial models, when we use the MSE to assess the predictive skill of the various models. However, an assessment of the upper tail, or the predicted probability of one or more claims per year, under the Brier score ranks the quarterly negative binomial models as the best model. As we are interested in correctly capturing the upper tail, we continue our analysis with the negative binomial model estimated on a quarterly basis with both topographical and climatological covariates. 

\begin{figure}[t]
	\centering
	\includegraphics[width=.8\linewidth]{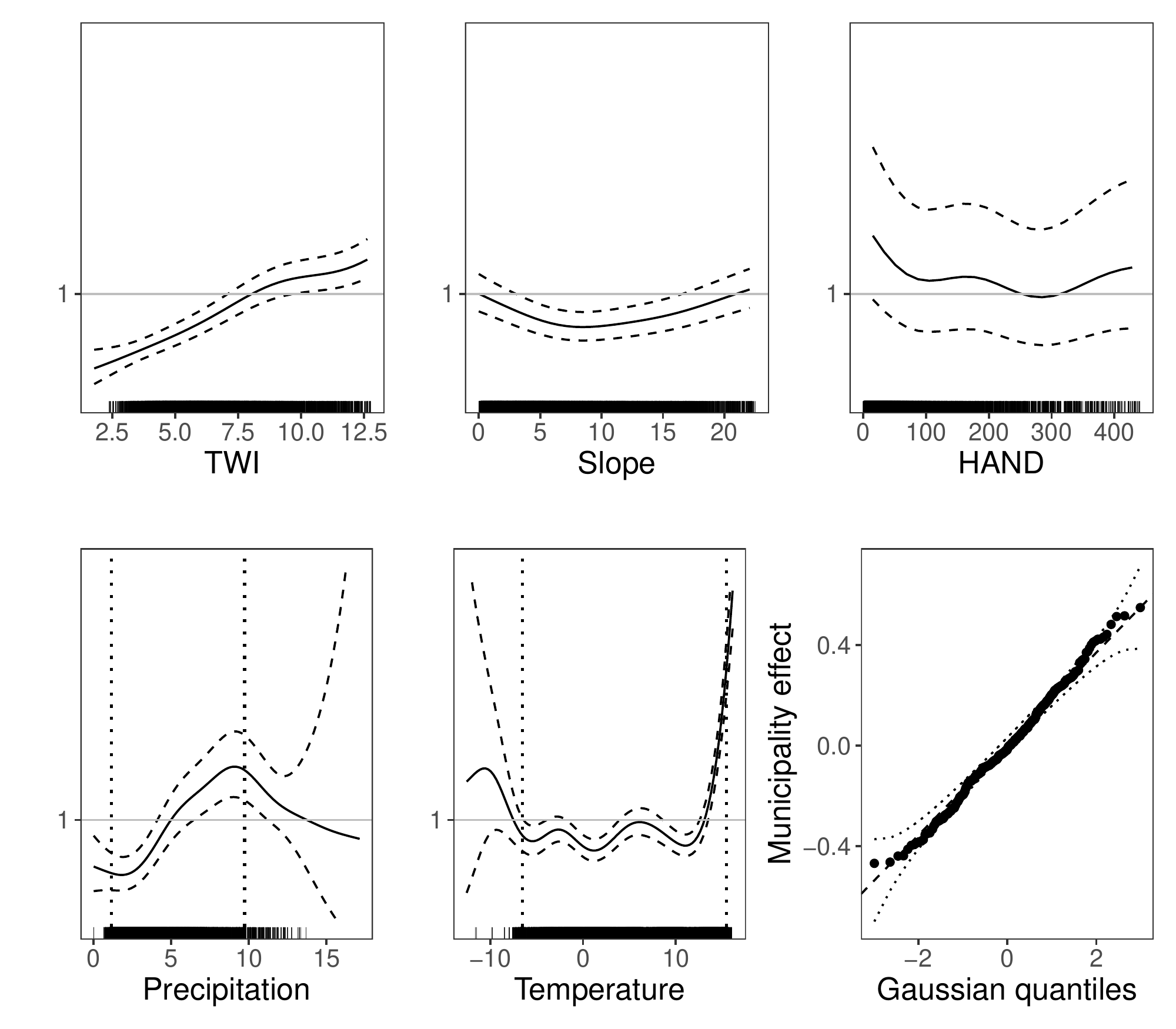}
	\caption{Multiplicative response effects of topographical (first row) and climatological (second row) covariates on water damage risk under the quarterly negative binomial model indicated by the mean effect (solid line) with a 95\% confidence band (dashed lines), all on a joint y-axis scale. 
	In-sample covariate values are indicated by black tick marks along the x-axes. For temperature and precipitation, dotted vertical lines show the 1st and 99th percentile, cf. Section \ref{sec:ClimateChangePrediction}. Bottom right: QQ-plot of the municipality random effect.}
	\label{fig:splines_ex_building}
\end{figure}

Figure \ref{fig:splines_ex_building} shows the estimated effect of the topographical and climatological covariates for the quarterly negative binomial model. Specifically, we show the multiplicative effect of these variables by taking the exponential of the additive effect. All response panels share the same scale for the y-axis, but the range is suppressed to prevent adaptation of the fitted model by direct competitors of Gjensidige. The top row shows the effect of the three topographical indices. The topographical wetness index (TWI) has a clear linear effect. The higher the wetness index the higher the expected number of claims. The effect of the slope index is shaped as a parabola with a minimum at approximately 10 degrees and a higher risk for both a lower and a higher slope. The effect of the slope index is difficult to interpret independently of the TWI which directly depends on the slope. The estimated spline for slope having a clearly pronounced shape gives evidence that information is added by considering both variables.
The HAND index has the largest effect when its value is between 0 and 100. This is intuitive since, at a certain point, it becomes irrelevant to move even higher above the nearest drainage point. The effect of precipitation is clearly non-linear; for low values the effect is flat, then close to linear between 3 and 9 mm of average daily precipitation in a quarter. The decreasing effect for precipitation values above 9 mm is somewhat counterintuitive and might be the result of little data in the upper tail. For temperature, the effect is small for average quarterly temperature of less than 10 degrees, while it is highly non-linear for higher values. See Section \ref{sec:discussion} for a further discussion of this effect. Lastly, we show a QQ-plot of the random effect of municipality which shows that the Gaussian assumption is fulfilled. 

\subsection{Risk assessment for a stationary climate}

\begin{figure}[!hbpt]
	\centering
	\includegraphics[width=.27\linewidth,height=.27\linewidth]{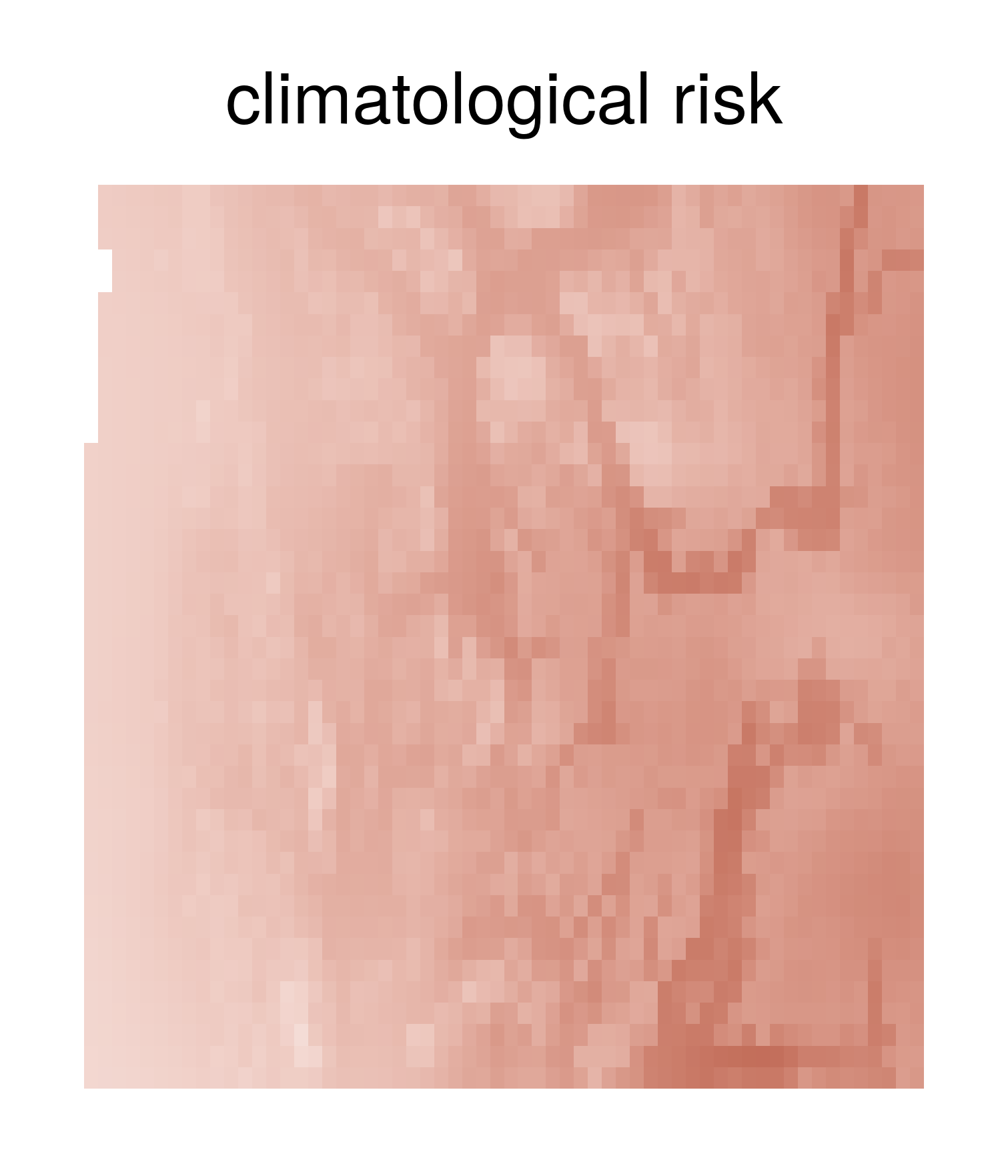}
	\includegraphics[width=.27\linewidth,height=.27\linewidth]{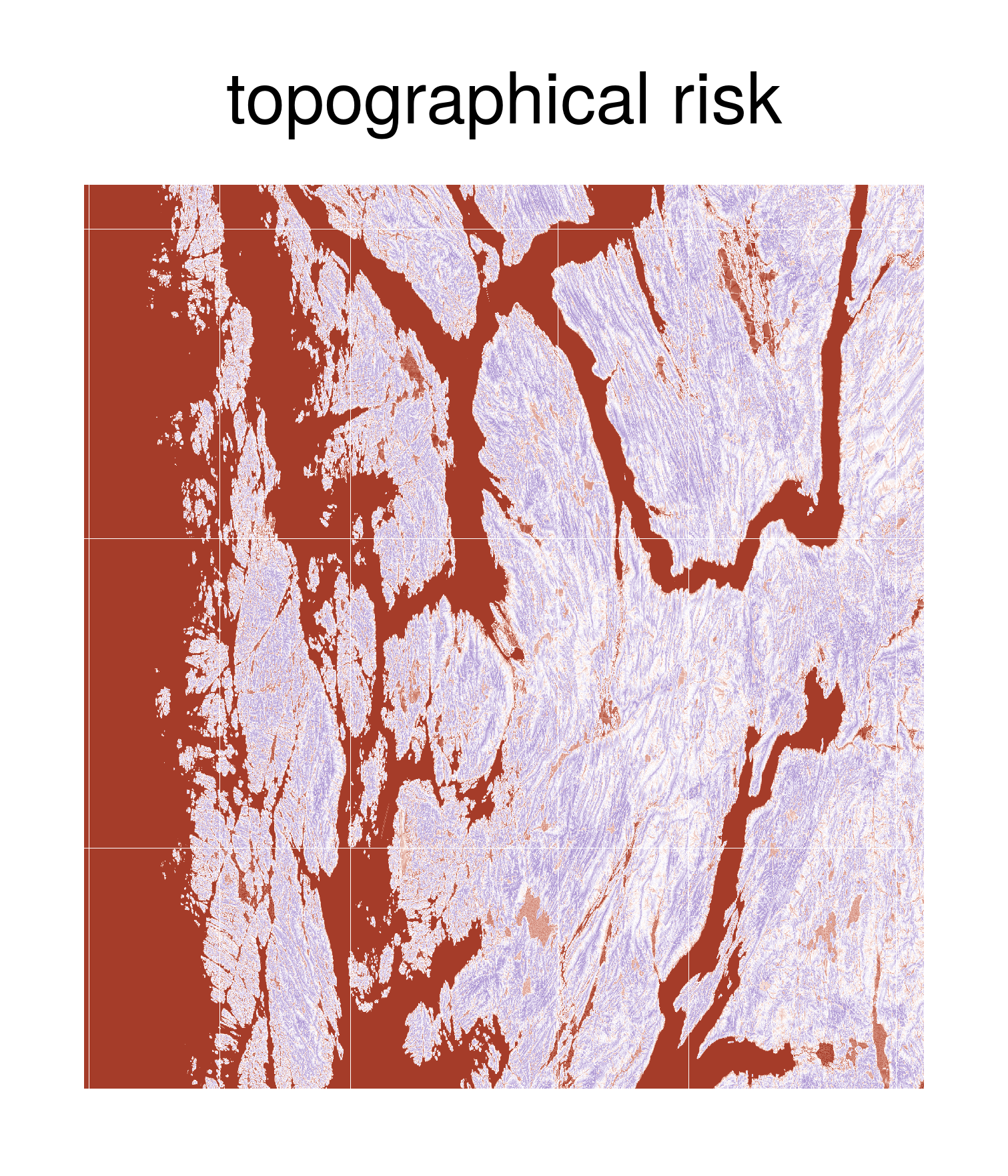}
	\includegraphics[width=.315\linewidth,height=.27\linewidth]{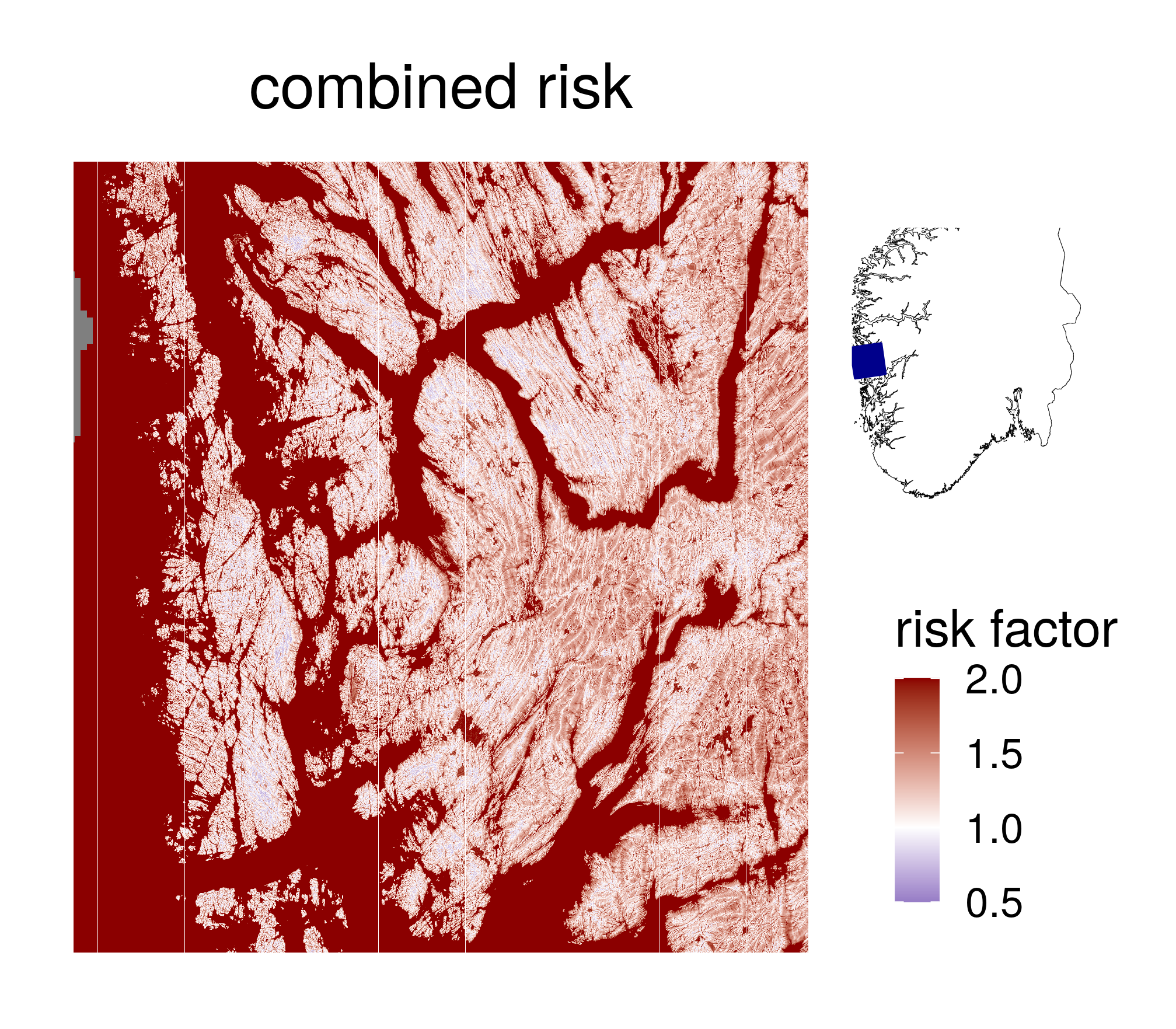}
	\includegraphics[width=.27\linewidth,height=.27\linewidth]{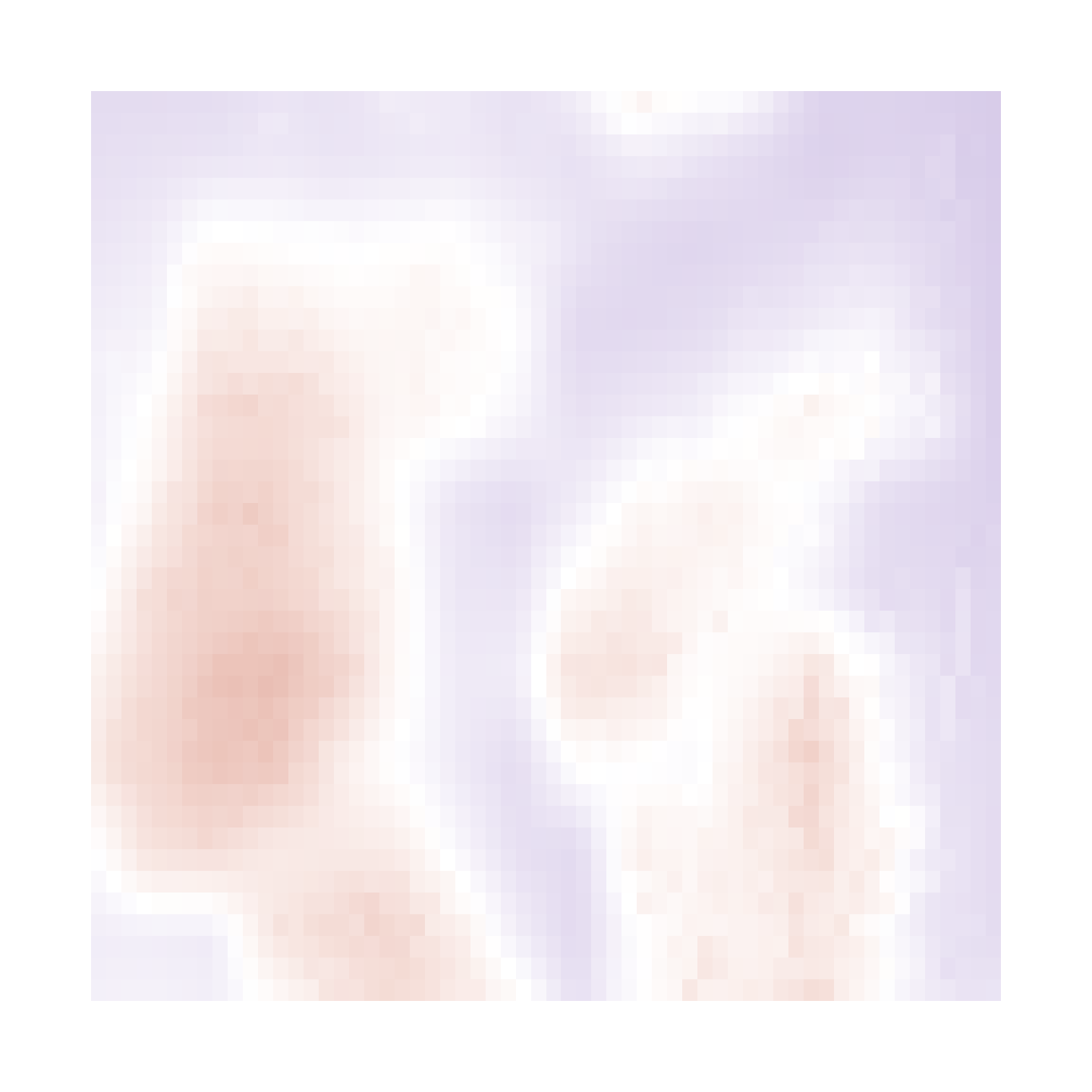}
	\includegraphics[width=.27\linewidth,height=.27\linewidth]{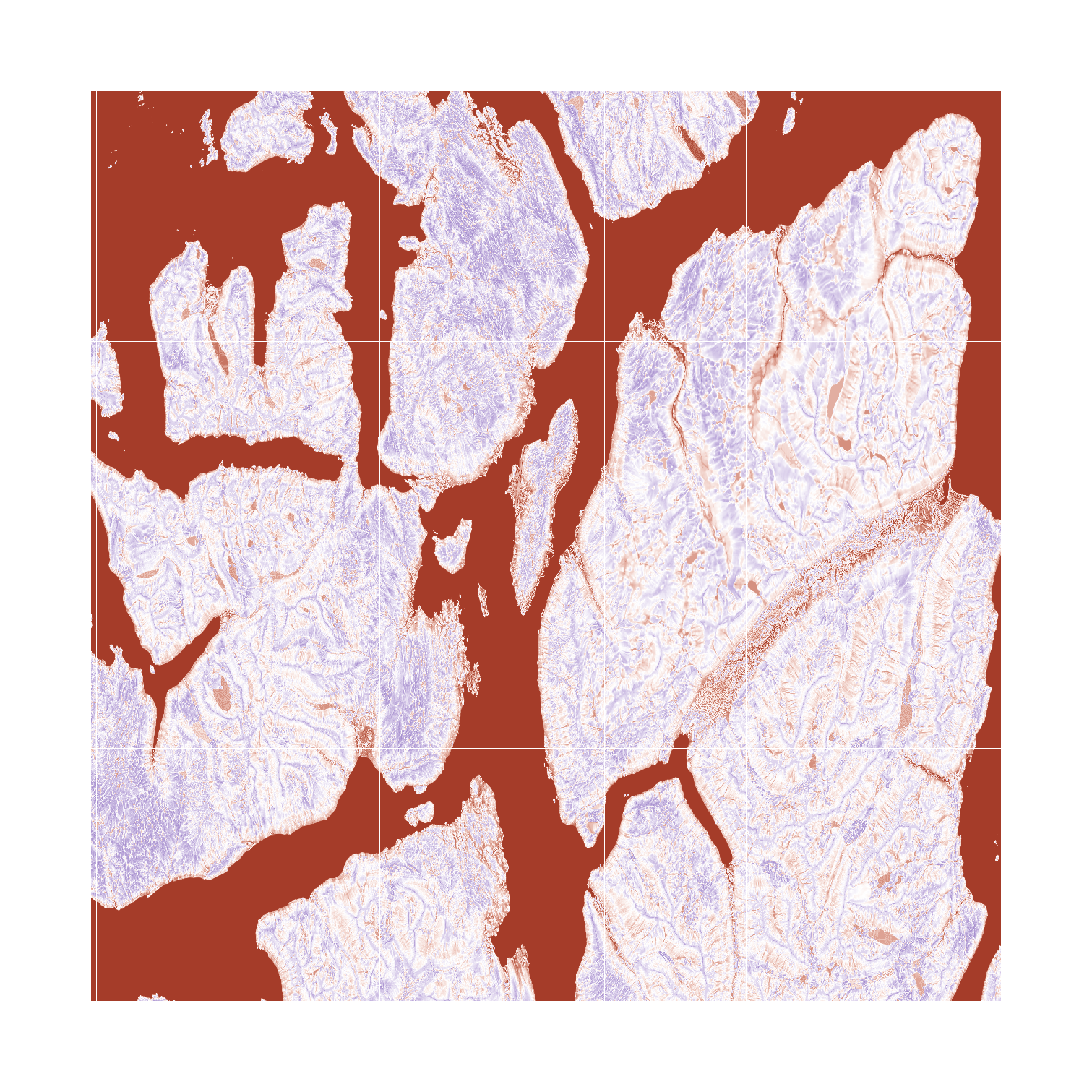}
	\includegraphics[width=.315\linewidth,height=.27\linewidth]{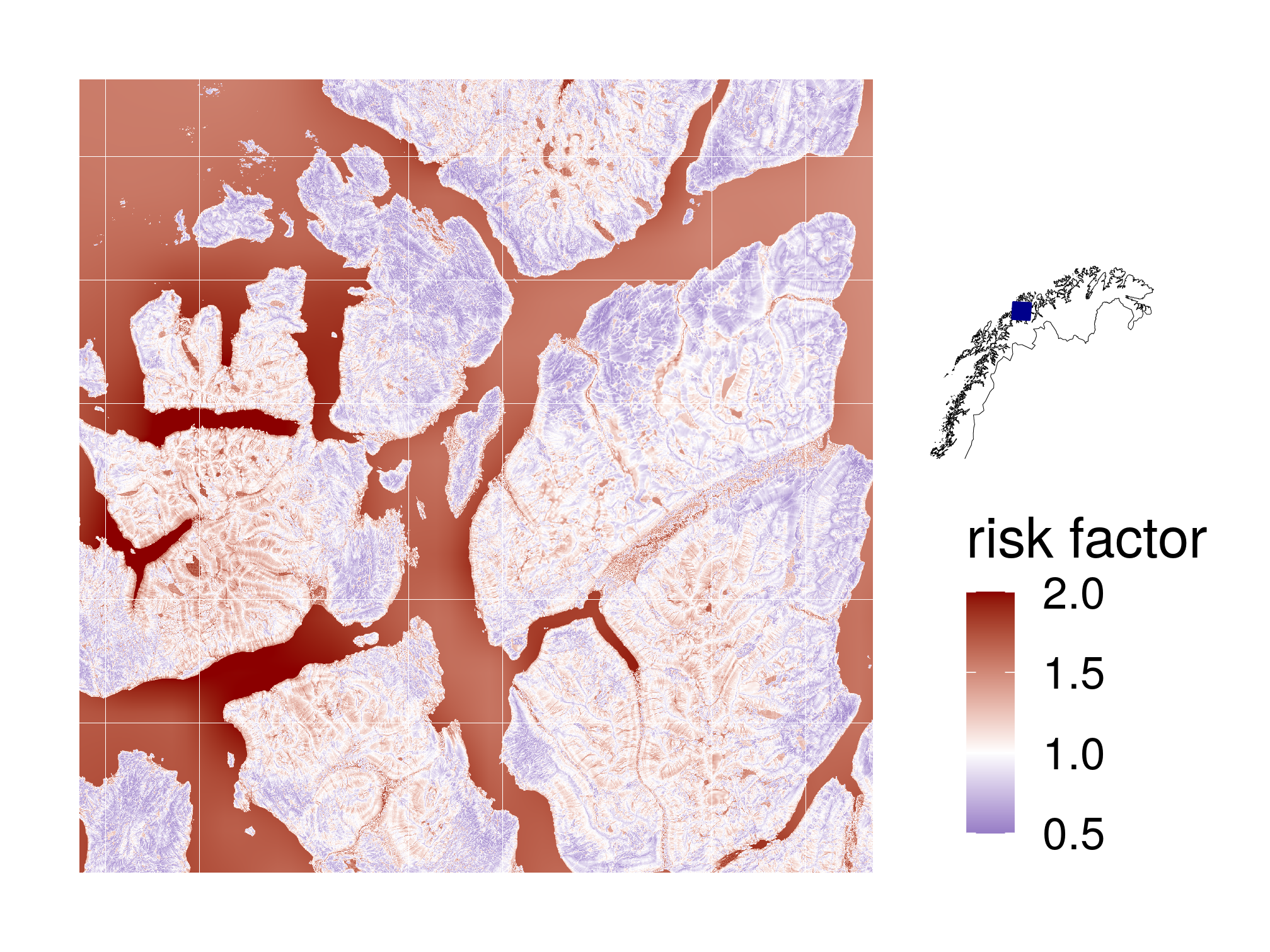}
		\caption{Climatological, topographical and combined risk factor for the greater Bergen area (top row) and greater Troms\o\ area (bottom row), see the small maps on the right hand side for the location of Bergen and Troms\o, respectively. The colorscale is centered at white, corresponding to the average risk, with below-average risk in blue and above-average risk in red.}
	\label{fig:risk_factors}
\end{figure}

As described in Section \ref{sec:risk_assessment}, we can decompose our model into several risk factors, corresponding to the groups of covariates. In particular, the model assigns a specific topographical risk $r_i^{\text{topo}}$ and a specific climatological risk $r_i^{\text{clim}}$ to each contract, see \eqref{eq:risk_factor}. These factors depend only on the location and not on any contract-specific characteristics. They can therefore be calculated for any location for which topographical/climatological data is available. Figure \ref{fig:risk_factors} shows maps of the corresponding sub-risks, as well as a combined map showing the product $r_i^{\text{topo}}r_i^{\text{clim}}$ for two example areas of $60 \times 60$ km, around the Norwegian cities of Bergen and Troms\o \, that obtain different risk profiles. The greater Bergen area, famous for heavy rainfall, gets assigned an above average (that is, $>1$) climatological risk, resulting in an above average combined risk. For Troms\o, the climatological risk and subsequently the combined risk is more varied within the considered region.

\subsection{Projected risk development due to climate change}\label{sec:ClimateChangePrediction}

\begin{figure}[!hbpt]
	\centering
	\includegraphics[width=.95\linewidth,height=.55\linewidth]{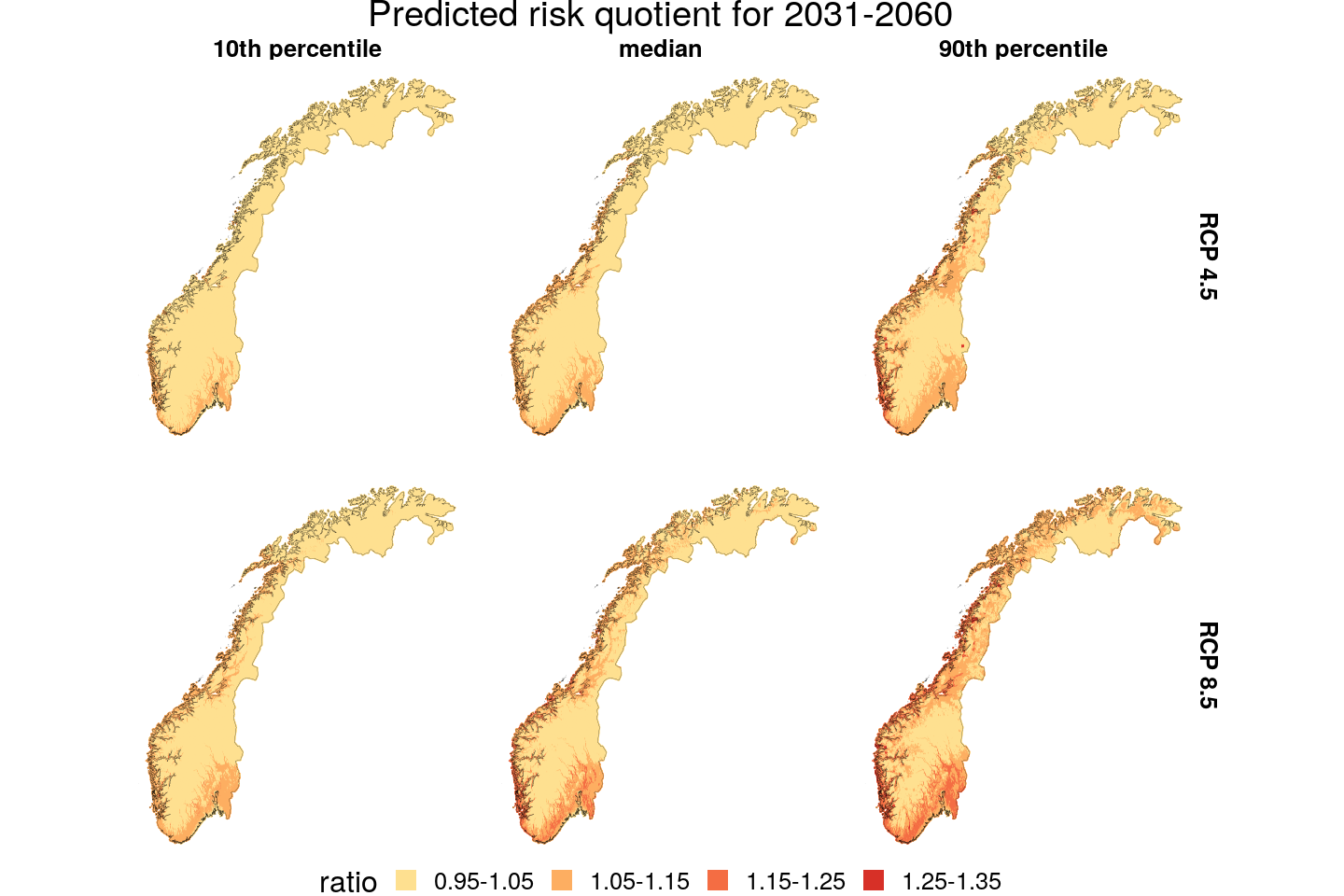}\\[1em]
	\includegraphics[width=.95\linewidth,height=.55\linewidth]{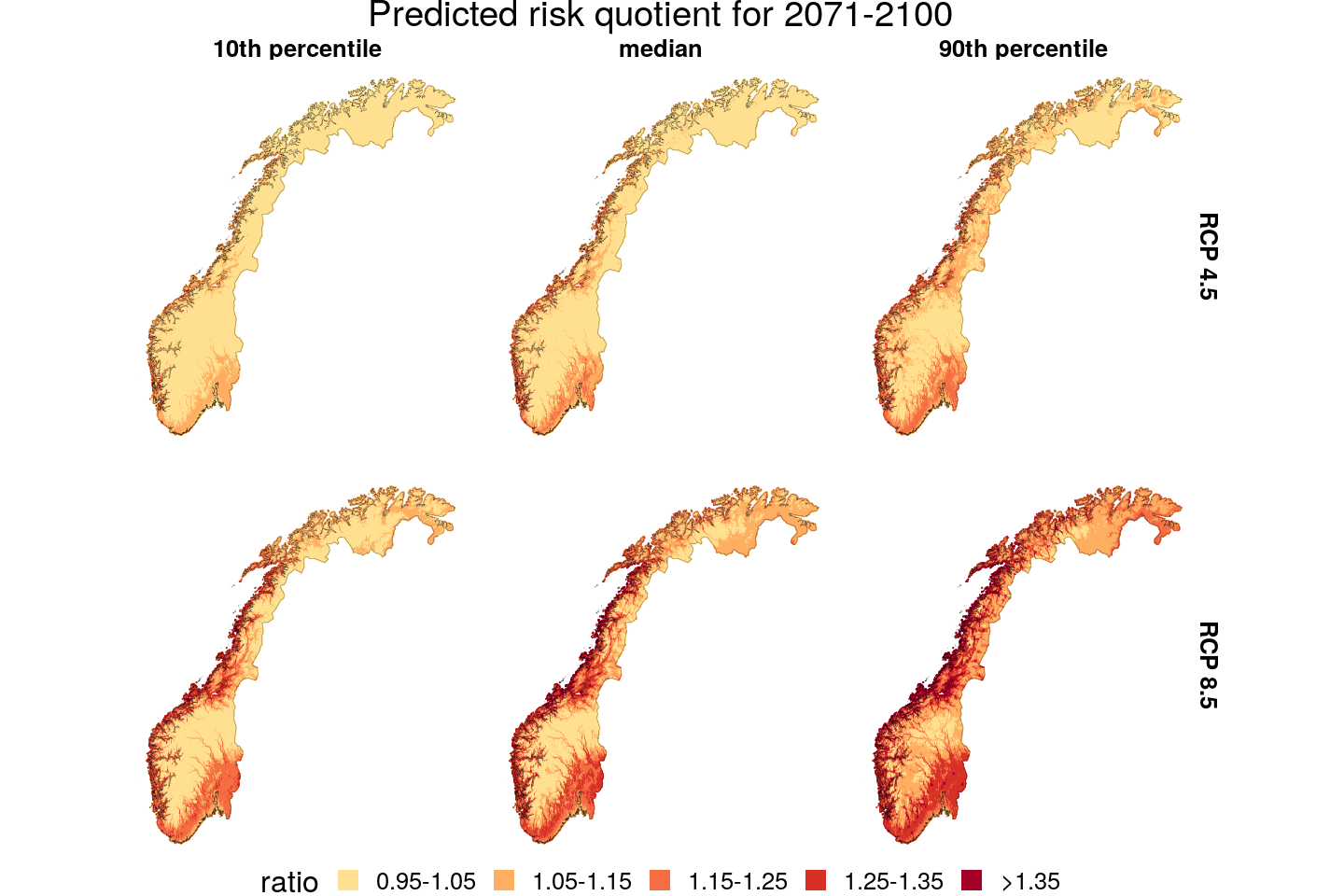}
	\caption{Ratio of projected future climatological risk for 2031-2060 (top) and 2071-2100 (bottom) and historical 1991-2020 climatological risk, see Section \ref{sec:risk_assessment} for a definition. Each row shows pointwise 10th, 50th and 90th percentiles of the probabilistic projections based on 12 climate models.}
	\label{fig:ratio_claims}
\end{figure}

\begin{figure}[!hbpt]
	\centering
	\includegraphics[width=\linewidth]{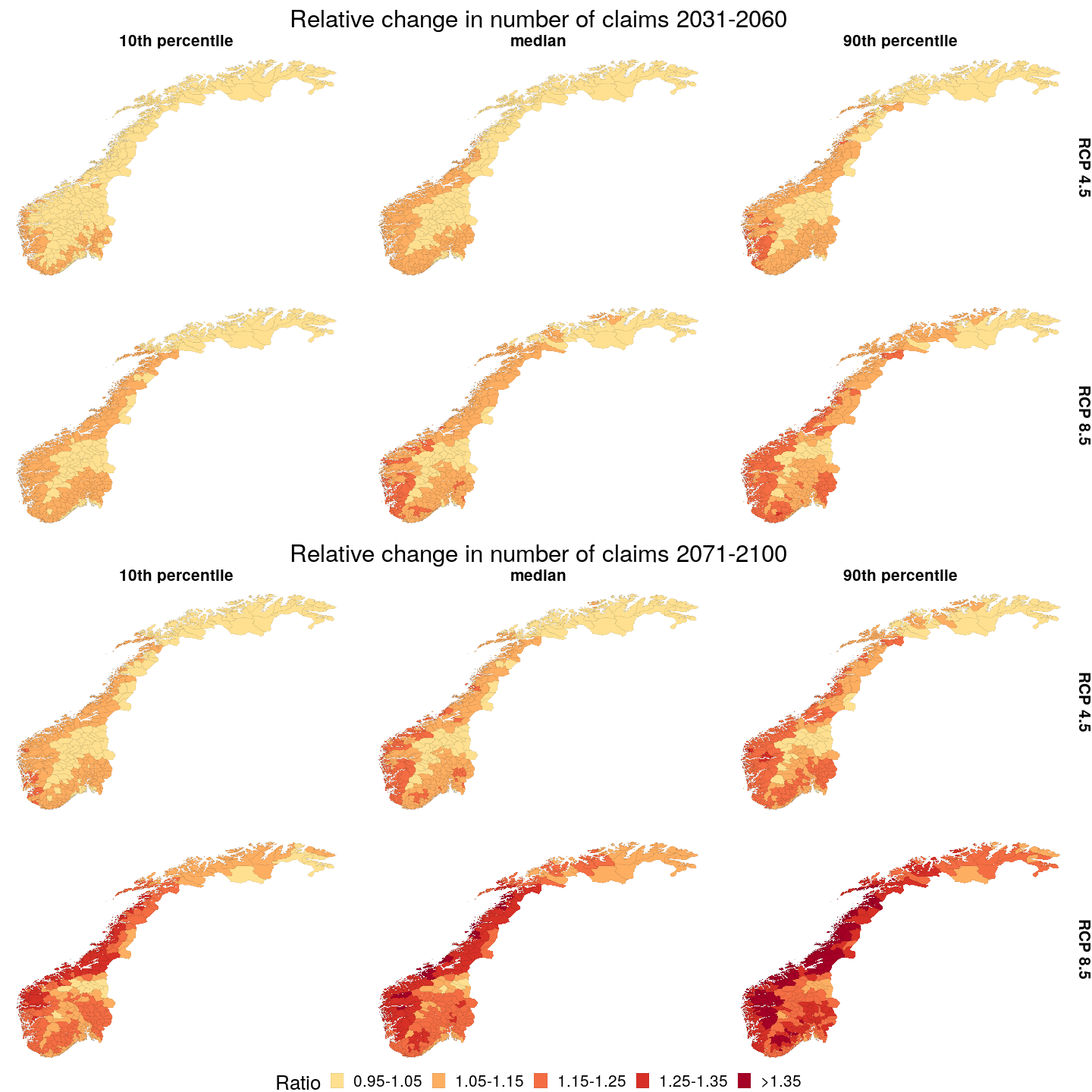}
	\caption{Projected relative change in number of claims for all municipalities in Norway based on climate projections for 2031-2060 (top) and 2071-2100 (bottom) relative to estimates for 2011-2020 based on 1991-2020 climatology. For each future period, the first row is based on climate projections under RCP 4.5 and the second row under RCP 8.5. The columns shows the 10th, 50th and the 90th percentile (for each municipality) based on 12 climate models.}
	\label{fig:hist_climate_ratio_claims}
\end{figure}

We employ 12 regional climate projections under both RCP 4.5 and 8.5 for the two future periods 2031-2060 and 2071-2100 in order to assess risk development due to climate change. It can be observed in Figure \ref{fig:splines_ex_building} that the estimated splines for the climatological variables exhibit implausible behavior towards the tail of the distribution of the training data.  For example, the spline for average daily precipitation decreases starting at 9.5 mm/day. Simultaneously, the uncertainty of the spline increases dramatically, suggesting that this effect likely constitutes an artifact of the statistical model rather than reflecting a causal relationship. This does not have much effect for the risk assessment in the current climate, since only few data points are located within the affected range. In climate projections for future time periods, however, the distribution of the climatological variables is shifted. As a result, the tail estimates of the splines can have much higher impact. To counteract an implausible extrapolation of the model beyond the range of the training data, we regularize the model by freezing the two climatological splines before the 1st percentile and past the 99th percentile of the training data, indicated as vertical dotted lines in Figure \ref{fig:splines_ex_building}. This extrapolation issue and associated model assumptions are further discussed in Section \ref{sec:discussion}.

To visualize future projected changes in risk, we consider the ratio of future and present climatological risk factors. Figure \ref{fig:ratio_claims} shows the 10th, 50th and 90th percentile of projected risk ratios based on the 12 different climate models. The projections show a significant increase in risk for the west coast up to and including the Lofoten archipelago, as well as for the southeastern part of the country around the capital, Oslo. Towards the end of the century, both RCP scenarios project large areas with increased risk of 25\% and higher. Under RCP 8.5 an increase of more than 35\% is projected for some areas. None of the considered scenarios project a decrease in risk for any part of the country.

The maps shown in Figure \ref{fig:ratio_claims} are exclusively based on the projected changes in mean quarterly precipitation and surface temperature. Alternatively, we can use our full model to project the change in number of claims per municipality. These projections incorporate topography and building-specific characteristics of the contracts within each municipality and thus more directly connect to Gjensidige's portfolio. The resulting projections are shown in Figure \ref{fig:hist_climate_ratio_claims}. These overall ratios are somewhat higher than for the climatological risk in Figure~\ref{fig:ratio_claims}, in that a larger overall area has a projected relative change of more than 15\% in Figure~\ref{fig:hist_climate_ratio_claims} compared to Figure~\ref{fig:ratio_claims}. 
This effect can partly be explained by the uneven spatial distribution of buildings within the considered municipalities: The majority of Norway's population lives in coastal areas, where a higher increase of risk is expected according to Figure~\ref{fig:ratio_claims}.

\begin{figure}[!ht]
 	\centering
 	\includegraphics[width=\linewidth]{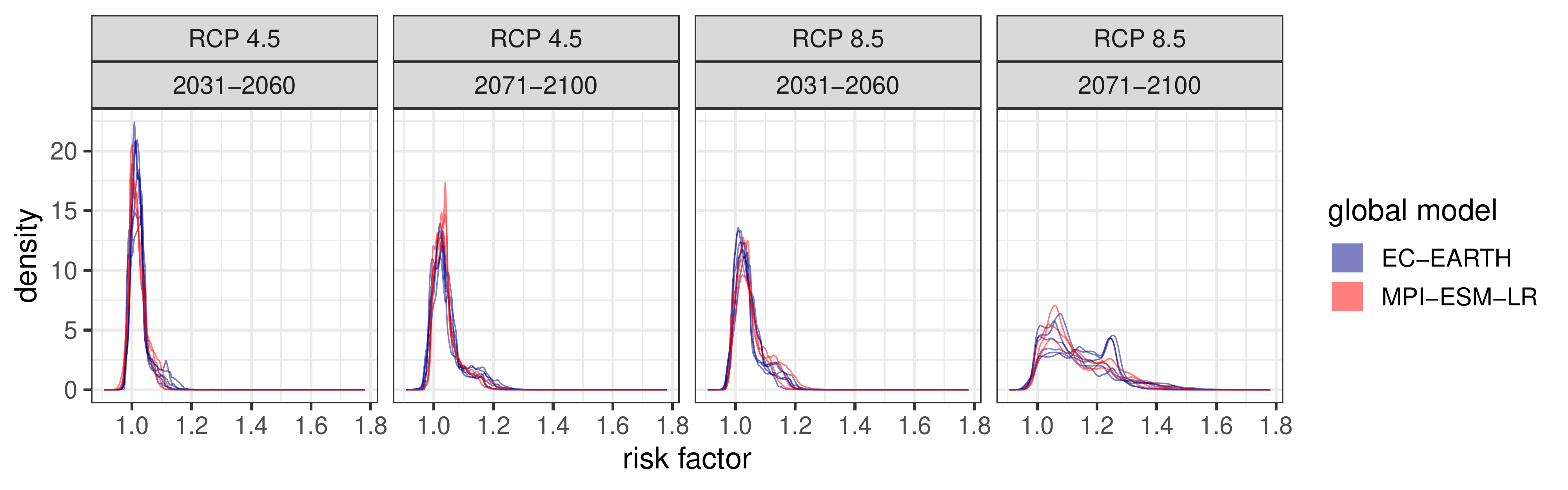}
 	\caption{Densities for relative risk factors derived from the climate projections compared to the reference period 1991-2020. Each density represents one climate model configuration (see Table \ref{tab:climate_data}), and shows the distribution of projected risk factors across all grid cells covering Norway. The colors of the densities highlight the two different driving GCMs.}
\label{fig:density_climate_models}
\end{figure}

Figure \ref{fig:density_climate_models} shows densities of projected relative climatological risk factors across all grid cells covering Norway for the different climate model configurations. The figure shows that these marginal distributions are rather similar between the different climate models for each projection period and RCP scenario. All densities for the future period 2031-2060 as well as densities for 2071-2100 under RCP 4.5 have a mode within the interval $[1,1.05]$ while the projections for 2071-2100 under RCP 8.5 are more diffuse with modes within the interval $[1,1.3]$. Furthermore, the densities are skewed with a heavy upper tail and an increasing skewness for a more pessimistic RCP and/or a time period further in the future.

\clearpage
\section{Discussion}\label{sec:discussion}

\begin{figure}[!ht]
	\centering
	\includegraphics[width=.95\linewidth]{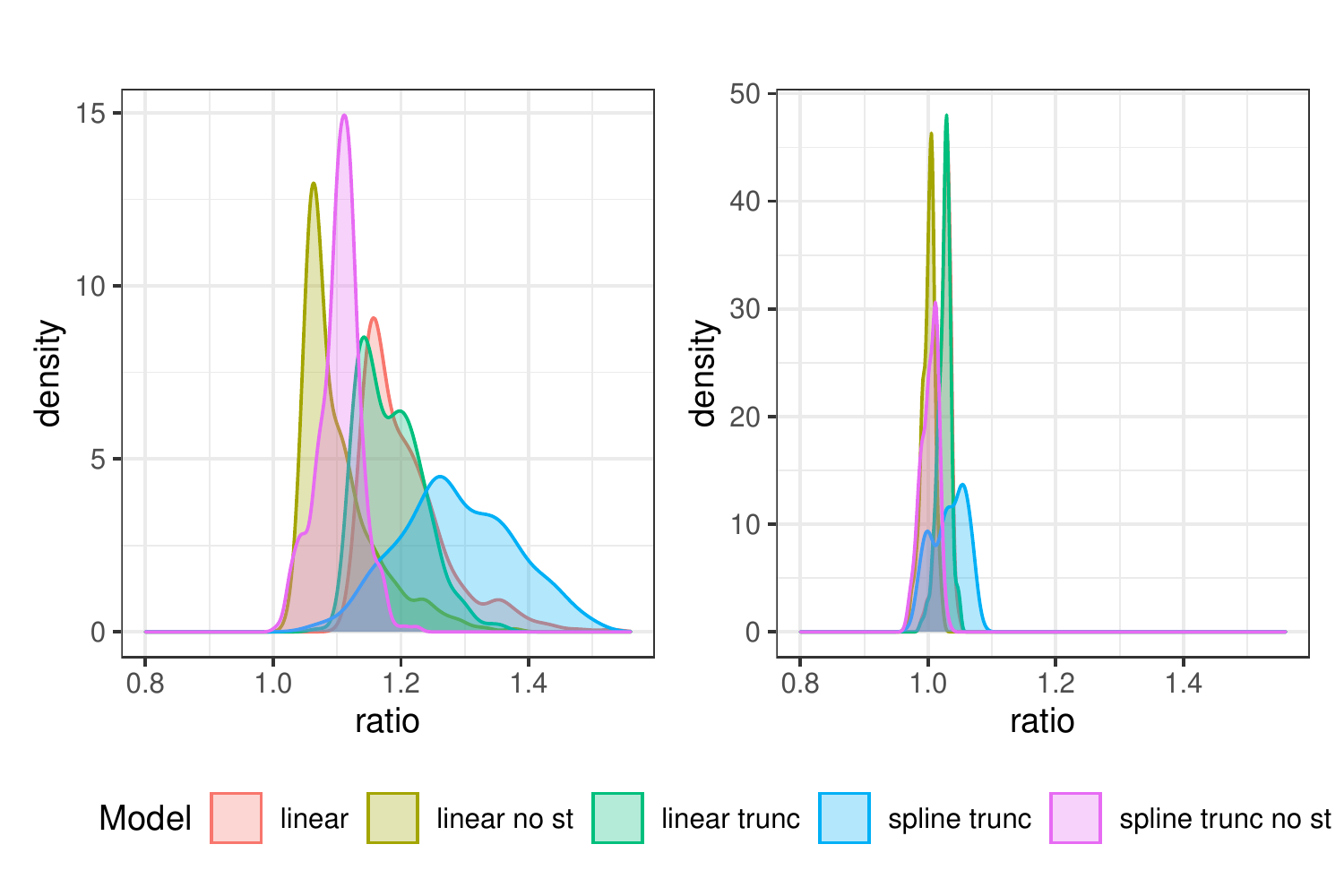}
	\caption{Left: Densities of 90th percentile of relative change in number of claims due to climate change under RCP 8.5 for the period 2071--2100 across all gridpoints in Norway. Right: Densities of 10th percentile of relative change in number of claims due to climate change under RCP 4.5 for the period 2031--2060 across all gridpoints in Norway. Each color represents projections from different regression models: \textit{spline} represents the model described in Section~\ref{sec:model}, \textit{linear} means that the effect of precipitation and surface temperature is linear on log scale, \textit{trunc} means precipitation and surface temperature is truncated at the 1st and 99th percentile of the training data and \textit{no st} means that surface temperature is excluded from the model.}
	\label{fig:extrapolation}
\end{figure}

The projection of water damage risk to a future climate involves extrapolating the model in \eqref{eq:logLik} beyond the range of the climatological indices for the reference period 1991-2020. As Norwegian climate is expected to get warmer and wetter \citep{hanssen&2009}, the extrapolation primarily involves an extrapolation beyond the upper limit of the in-sample values, see Figure~\ref{fig:splines_ex_building}. The projections presented in Section~\ref{sec:ClimateChangePrediction} are based on splines truncated at the 1st and 99th percentile of the in-sample distribution of covariate values. In order to assess the effect of this modeling choice, we have also formed water damage risk projections using a linear model for the climatological covariate effect applied with or without truncation, as well as models that only use a precipitation index rather than both precipitation and temperature indices. A model comparison is shown in Figure~\ref{fig:extrapolation}. Figure~\ref{fig:extrapolation} shows that the modeling choice presented in Section~\ref{sec:ClimateChangePrediction} is the least conservative of the considered options. In particular, we see that the projected increase in water damage risk is, to a large extent, driven by the inclusion of a temperature index. However, a comparison of the predictive performance of these models in the current climate through a cross-validation study as presented in Section~\ref{sec:model selection} shows that the non-linear splines with both temperature- and precipitation-based indices significantly outperforms the other modeling options (results not shown).  

In Figure~\ref{fig:splines_ex_building}, we see a steep increase in the spline for high temperatures. Physically, there is no immediate explanation why high temperatures should generate more claims. Thus these temperatures most likely act as a proxy for some other phenomenon expressed in the data. One hypothesis would be that high surface temperatures are associated with more frequent cloud bursts typically seen during the warm season and in particular on hot days (local, convective precipitation). In urban areas such events are known to bring potential risk of vast water damage due to a high proportion of non-porous surfaces, often in combination with limited capacity of the sewer system. Claims generated from such conditions will possibly ascribe high values of the temperature variable as the key source for their occurrence. 
Rural areas also experience hot days with cloud bursts, on average to the same extent as urban areas. Buildings in rural areas are less vulnerable to these events and, for these areas, we would expect the spline to have less of a steep increase for high temperatures. However, as the majority of the buildings in the insurance portfolio are situated in urban areas the model estimates are formed primarily from the pattern of urban claims.

\section{Conclusions}

This paper proposes a nationwide model for risk of rainfall-induced water damage in Norway. The model incorporates local topographical and climatological information as well as contract-specific property data. The overall risk assessment can be decomposed into several factors which, in particular, allows for the derivation of topographical and climatological risk maps covering the entire country. Climate projections yield projections for changes in water damage risk in future decades due to climate change. Based on a multi-model ensemble of regional climate projections, we project a spatially varying increase in risk of as much as 25\% for large areas towards the end of this century.

Insurance companies' risk models materialize in tariffs that price the individual customer and the individual object. Historical damages and losses---including those caused by pluvial flooding---form the basis for the tariffs, and trends together with knowledge of future changes make the tariffs predictive. The results of this study improve the pricing quality in multiple ways. Gjensidige will gain knowledge about future developments in the overall level of losses, the geographical loss distribution related to present and future climate and the risk level of individual buildings due to their location in the terrain. These are aspects that expand and challenge current risk assessment practice in the insurance sector. The model provides not only precise risk information for Gjensidige's existing customers, but also any potential new object---built or planned---in Gjensidige's future portfolio. All of this is crucial information for pricing according to risk at both portfolio and customer level.

Risk associated with climate and climate change plays a special role compared to other risks due to global efforts for climate change adaptation and mitigation. The risk models that have been developed here provide Gjensidige with a basis for redistribution of premiums at building level (topography) and future redistribution and adaptation at geographical level. More importantly, the new knowledge can be utilized to reduce risk for both municipalities and individual customers, supporting and furthering a sustainable business strategy. Municipalities can gain valuable insights regarding investment priorities such as upgrades of sewerage networks or planning of new residential or commercial areas. Gjensidige will also seek to use the results to develop preventive measures that customers can implement themselves.

After an earlier trial period, Norwegian insurance companies will from 2022 provide claims data related to weather and climate to the Norwegian authorities. The purpose of this program is for the municipalities to get an overview of areas with high risk exposure so that measures can be taken where they are expected to have the greatest effect. The results of the current study provide valuable knowledge regarding expected future changes in this context. Models of the type presented here can help ensure that measures are prioritized to provide the greatest possible societal benefit, both by preventing economic and social costs and because every damage entails a CO$_2$ cost. The most effective climate change adaptation and mitigation strategy for an insurance company is to prevent damage occurrence.

\section*{Acknowledgments}
We thank Andreas Lura for extracting the insurance data, Magne Aldrin for helpful discussions and Annabelle Redelmeier for help with the manuscript. 



\end{document}